\documentclass[%
 reprint,
 longbibliography,
superscriptaddress,
 amsmath,amssymb,
 aps,
pra,
]{revtex4-1}
\usepackage{graphicx}
\usepackage{dcolumn}
\usepackage{bm}
\usepackage{subfigure}  

\begin{document}

\preprint{APS/123-QED}

\title{Non-equilibrium Fractional Hall Response After a Topological Quench}

\author{F. Nur \"{U}nal}
\email{fatmanur@bilkent.edu.tr}%
\affiliation{Laboratory of Atomic and Solid State Physics, Cornell University, Ithaca, New York 14853, USA}
\affiliation{Department of Physics, Bilkent University, Ankara, Turkey}

\author{Erich J. Mueller}
\affiliation{Laboratory of Atomic and Solid State Physics, Cornell University, Ithaca, New York 14853, USA}

\author{M. \"{O}. Oktel}
\affiliation{Department of Physics, Bilkent University, 06800 Ankara, Turkey}




\date{\today}

\begin{abstract}
We theoretically study the Hall response of a lattice system following a quench where the topology of a filled band is suddenly changed. In the limit where the physics is dominated by a single Dirac cone, we find that the change in the Hall conductivity is two-thirds of the quantum of conductivity. We explore this universal behavior in the Haldane model, and discuss cold-atom experiments for its observation. Beyond linear response, the Hall effect crosses over from fractional to integer values. We investigate finite-size effects, and the role of the harmonic confinement.

\end{abstract}

\maketitle

\section{INTRODUCTION}
What happens when a system is suddenly driven between two topologically different phases? Several authors have tried to answer this question in the context of Chern insulators \cite{Rigol,CooperPRL,KehreinPRB,Sengstock,Refael,DynBuildup,Cooper2}. Aspects of the original topology survive the quench, but most physical observables (edge currents, Hall conductivity) appear to be non-universal. Here, we study the non-equilibrium Hall response following a quench where the mass term of a single Dirac cone changes sign, and apply these results to understanding quenches in Chern insulators. Such non-equilibrium questions are at the forefront of research into topological systems \cite{Yuzbashyan,Rigol,CooperPRL,Cooper2,Sengstock,Refael,KehreinPRB,DynBuildup,AditiPRB}. We find that for symmetric quenches dominated by a single Dirac cone, the Hall conductivity universally changes by $2e^2/3h$, which is two-thirds of the quantum of conductivity.

We argue that this non-equilibrium topological response can be observed in cold-atom experiments. Jotzu et al. \cite{HaldaneExp} recently implemented the Haldane model \cite{Haldane} using fermionic potassium atoms, and observed transitions between ordinary insulators (OIs) and Chern insulators (CIs). They are capable of quenching between these phases. Here, we analyze these transitions, and discuss a protocol for observing the non-equilibrium fractional Hall response.

To establish our central result, we first analytically calculate the non-equilibrium response for a single Dirac cone in a two-band model. We then show that the universal fractional Hall response of a single Dirac cone appears in quenches of the Haldane model. We consider both an infinite geometry as well as a strip configuration. The latter configuration lets us study the evolution of the edges modes. In addition to linear response, we explore the full time dynamics of this system in an electric field. We find that the currents perpendicular to the field are time-dependent. At short times, they are consistent with the fractional Hall conductivity predicted by our linear response theory. At long times, they instead correspond to an integer Hall response.

One important question for observing these effects is the role of finite system size. We find that for narrow strips the equilibrium Hall response deviates from the integral value predicted by bulk calculations. The deviation vanishes as the inverse of the system size. A related question is the role of harmonic confinement. Remarkably, we find that even though the edges of a harmonically trapped cloud are metallic, one can observe a quantized Hall response. Furthermore, we show that by analyzing currents inside the trapped cloud, one can infer the Hall conductance, without applying an electric field. When discussing the neutral atomic systems we will continue to use the language of electrodynamics: The electric field is a force, current is particle current, and the charge is simply unity.

The paper is organized as follows: In the next section, we calculate the Hall response of a single Dirac cone following a quench that inverts its mass sign. Section \ref{sec Haldane} generalizes this discussion to the Haldane model. In Sec.\ref{sec linear response}, we analyze the nonlinear characteristics of the out-of-equilibrium Hall response. Then in Sec.\ref{sec strip}, we consider a strip configuration and investigate finite size effects. Section \ref{sec harmonic trap} focuses on the effect of a harmonic confinement on the Hall conductivity of a strip, and discusses correspondences to cold atom experiments. We conclude our discussion with Sec.\ref{sec conclusion}.

\section{SINGLE DIRAC CONE} \label{single dirac section}
Near a topological transition of a $2D$ system, the low-energy physics is described by a massive Dirac model \cite{XiaoRev,Haldane, HasanKaneRev}. We parameterize the static Hamiltonian by two quantities; the gap $\Delta$ and the `speed of light' $c$,
\begin{equation}
{\cal H}(\vec{k})= \begin{pmatrix}
\Delta & cke^{i\theta}\\\\
cke^{-i\theta} & -\Delta
\end{pmatrix},
\end{equation}
where $ke^{i\theta}=k_x+ik_y$ expresses the $2D$ momentum as a complex number. When $\Delta=0$, the spectrum consists of a gapless Dirac cone at $\vec{k}=0$. For nonzero $\Delta$, a gap with magnitude $2|\Delta|$ opens at zero momentum. The energies $\varepsilon_{n}=(-1)^n\sqrt{\Delta^2+c^2k^2}$ for the band index $n=1,2$ are independent of the sign of the gap (mass). The corresponding eigenstates are
\begin{multline} \label{eq Dirac eigen st.s}
u_{n}(\vec{k})= \frac{1}{N_n(\vec{k})}
\begin{pmatrix}
(-1)^ne^{i\theta}ck/\Delta\\\\
\sqrt{1+\left(\frac{ck}{\Delta}\right)^2}-(-1)^n
\end{pmatrix},
\end{multline}
where $N_n(\vec{k})=\sqrt{2}\sqrt{1+\left(\frac{ck}{\Delta}\right)^2-(-1)^n\sqrt{1+\left(\frac{ck}{\Delta}\right)^2}}$ is the normalization. One imagines that the negative energy modes are filled, and the positive energy modes are empty. The contribution to the Hall conductivity from these low-energy modes can be calculated from the Chern number $C$ via $\sigma_H=Ce^2/h$ \cite{TKNN}. The Chern number is given by
\begin{equation}
C_n=\frac{1}{2\pi}\int d^2k\;\Omega_n(\vec{k}),
\end{equation}
where the Berry curvature is
\begin{equation} \label{Berry curv definition}
\Omega_{n}(\vec{k})=-i\langle\partial_{\vec{k}}u_n(\vec{k})|\times|\partial_{\vec{k}}u_n(\vec{k})\rangle =\frac{(-1)^{n+1}c^2\Delta}{2(\Delta^2+c^2k^2)^{3/2}}.
\end{equation}

The Berry curvature is odd in the gap parameter, and the Chern numbers of the cones $C=\pm1/2$ depend on the sign of $\Delta$, even though the energy spectrum does not. The Chern number for a given band must be an integer, so the modes neglected in the low-energy description must also supply a half-integer Chern number. We are envisioning a quench which does not affect these higher energy modes, and their contribution to the Hall conductivity will not change during the quench.

We imagine suddenly flipping the sign of the $\Delta$. The probability $P_n(\vec{k})$ of a particle of momentum $k$ being found in the $n^{th}$ band following this quench is simply the inner product between $u_1(\vec{k})$, and $\bar{u}_n(\vec{k})$, where $\bar{u}_n$ is given by Eq.(\ref{eq Dirac eigen st.s}) with $\Delta\rightarrow-\Delta$. This yields
\begin{equation} \label{P1, P2 after quench}
P_1(\vec{k})=\frac{c^2k^2}{\Delta^2+c^2k^2},\quad  P_2(\vec{k})=\frac{\Delta^2}{\Delta^2+c^2k^2}.
\end{equation}

After the quench, the system is in a non-equilibrium state. Hence, each $k$ point contributes to the Hall conductivity with its Berry curvature weighted by its occupation probability \cite{Cooper2,KehreinPRB},
\begin{equation} \label{sigma_H,neq def.}
\sigma_{H, neq}=\frac{e^2}{h}C_{neq}=\frac{e^2}{2\pi h}\sum_n\int d^2k P_n(\vec{k}) \Omega_n(\vec{k}).
\end{equation}
Any effects from coherences between the bands average out over a time $\tau\sim h/\Delta$, and can be neglected in linear response. Eq.(\ref{sigma_H,neq def.}) reduces to the usual TKNN formula \cite{TKNN} for the ground state where the occupation probability of the lowest band is one at each $k$ and the higher bands are completely empty. For partially filled bands, the dimensionless Hall conductivity $C_{neq}$ is clearly not quantized and can take any value. However, in a mass-sign-inverting quench, the symmetry between initial and final states will yield a universal result. In particular for our case, the integrals in (\ref{sigma_H,neq def.}) are elementary and
\begin{equation} \label{sigma_H def.}
C_{neq}=\frac{1}{3}-\frac{1}{6}=\frac{1}{6}.
\end{equation}
Here, the first (second) term is the contribution coming from the lower (upper) band. This fractional Hall response is independent of $\Delta$. Although the Berry curvatures and the number of excited particles are highly sensitive to the value of the gap, they remarkably balance each other in symmetric quenches and one always ends up with $1/6$-Hall response. If the quench is performed in the opposite direction, $C_{neq}$ changes sign to $-1/6$. Since the initial Chern number was $\pm1/2$, the change in the Hall conductivity in such symmetric quenches is always $\pm2/3$.

\begin{figure}
\centering
\includegraphics[width=0.47\textwidth]{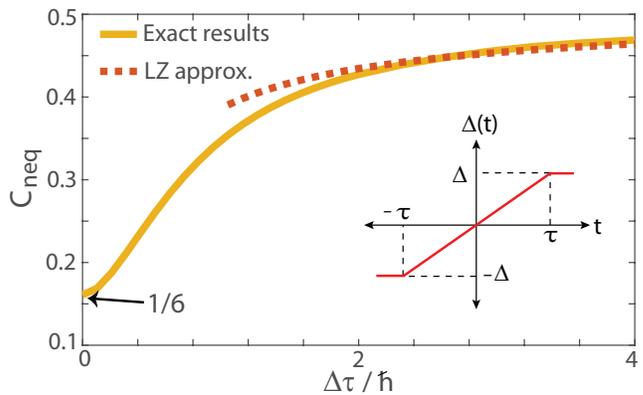}
\caption{ Non-equilibrium Hall conductivity, $C_{neq}$, of a single Dirac cone following a ramp of the gap from $-\Delta$ to $\Delta$ in time $2\tau$ as shown in inset. $C_{neq}$ is measured in units of $e^2/h$, and the product $\Delta\tau/\hbar$ is dimensionless. In the sudden quench limit ($\tau=0$), $C_{neq}=1/6$ and as $\Delta\tau\rightarrow\infty$, $C_{neq}\rightarrow1/2$. The dashed line shows prediction of the LZ theory from Eq.(\ref{eq LZ}).  } \label{fig1}
\end{figure}

For completeness, we also consider sweeps with finite rate where the gap is linearly ramped from $-\Delta$ to $\Delta$ within time $2\tau$ as illustrated in the inset of Fig.\ref{fig1}. We solve the Schr\"{o}dinger equation
\begin{equation}
i\hbar\frac{\partial}{\partial t} |\Psi(\vec{k},t)\rangle={\cal H}(\vec{k},t)|\Psi(\vec{k},t)\rangle,
\end{equation}
assuming the lower band is initially full. The adiabatic eigenstates of the Hamiltonian traverse an avoided crossing as the sign of the gap changes. At each $k$, there is an independent Landau-Zener (LZ) problem, with adiabaticity parameter $\eta_k=c^2k^2\tau/\hbar\Delta$. When $\eta_k\gg1$, the particle remains in the lower band, while $\eta_k\ll1$ it transitions with the probabilities in Eq.(\ref{P1, P2 after quench}). Adiabaticity always holds for modes with sufficiently large $k$. If the gap is large, $\Delta\gg\hbar/\tau$, then the majority of the modes which contribute to the Hall conductance will be in that region. One can then use the LZ approximation, $P_2(\vec{k},\tau)=e^{-\pi c^2 k^2\tau/\hbar\Delta}$ which yields a Hall conductivity
\begin{equation} \label{eq LZ}
C_{neq}^{LZ}=-\frac{1}{2}+\pi\sqrt{\frac{\Delta\tau}{\hbar}}e^{\pi\Delta\tau/\hbar} \text{erfc} \left(\sqrt{\pi\frac{\Delta\tau}{\hbar} }\right).
\end{equation}
Here, erfc is the complementary error function. For a general value of $\Delta\tau/\hbar$, we numerically calculate the excitation probabilities and the Hall conductivity (see Fig.\ref{fig1}). We see that for slow sweeps, the Hall conductivity saturates at $C_{neq}=1/2$, while rapid quenches yield $C_{neq}=1/6$. For $\Delta\tau/\hbar>2$, the Hall conductivity is well approximated by the LZ result Eq.(\ref{eq LZ}).

Our result, that the contribution to the non-equilibrium Hall conductivity from a single Dirac cone is $1/6$ of the quantum of conductivity, implies that one will see this same fraction in small quenches of any topological lattice system. To illustrate this universality, we study the Haldane model.

\section{THE HALDANE MODEL} \label{sec Haldane}
\subsection{Infinite System} \label{sec Haldane infinite}
\begin{figure}
\centering
\subfigure{\includegraphics[width=0.15\textwidth]{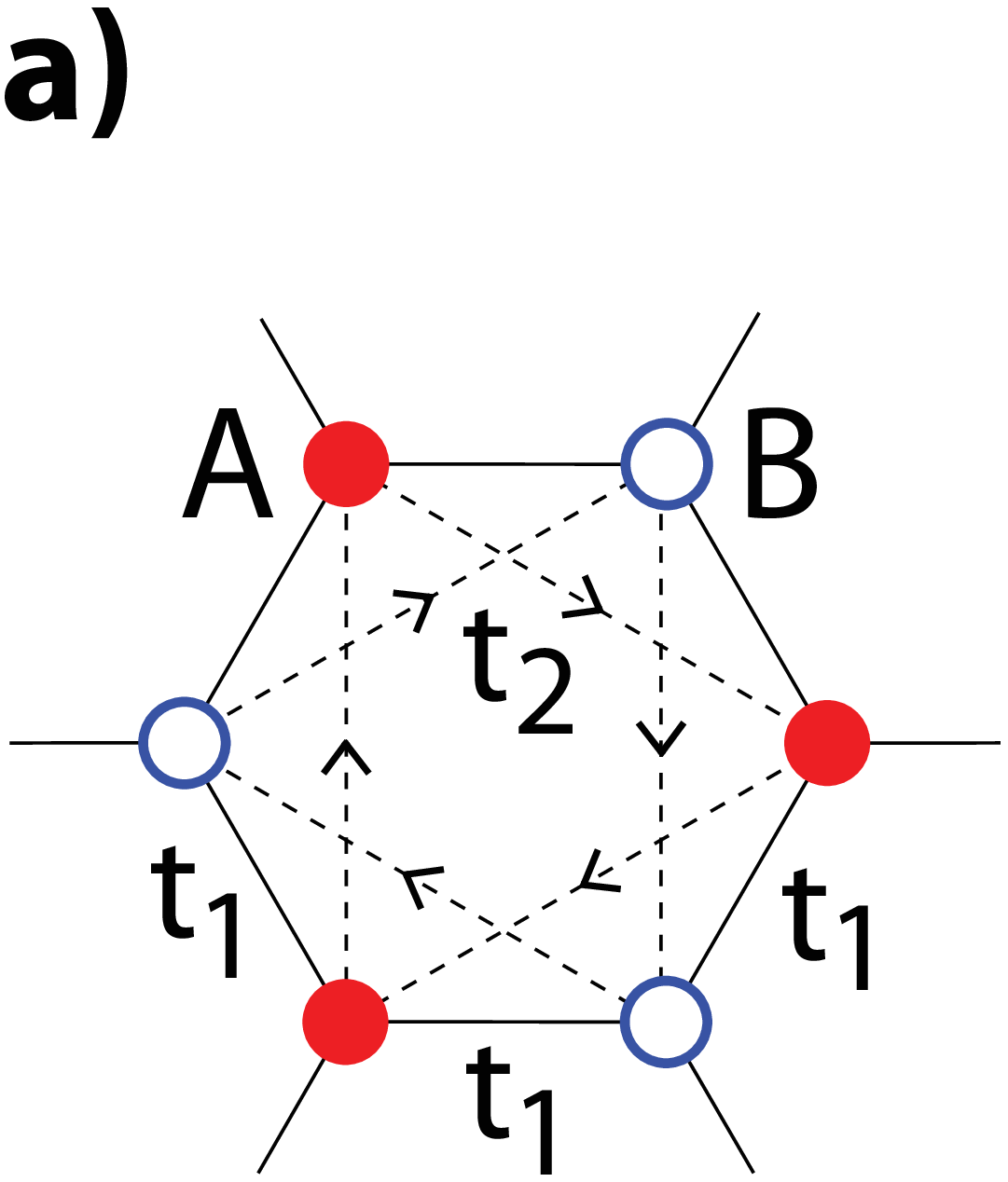} \label{fig Haldane lattice}}
\subfigure{\includegraphics[width=0.31\textwidth]{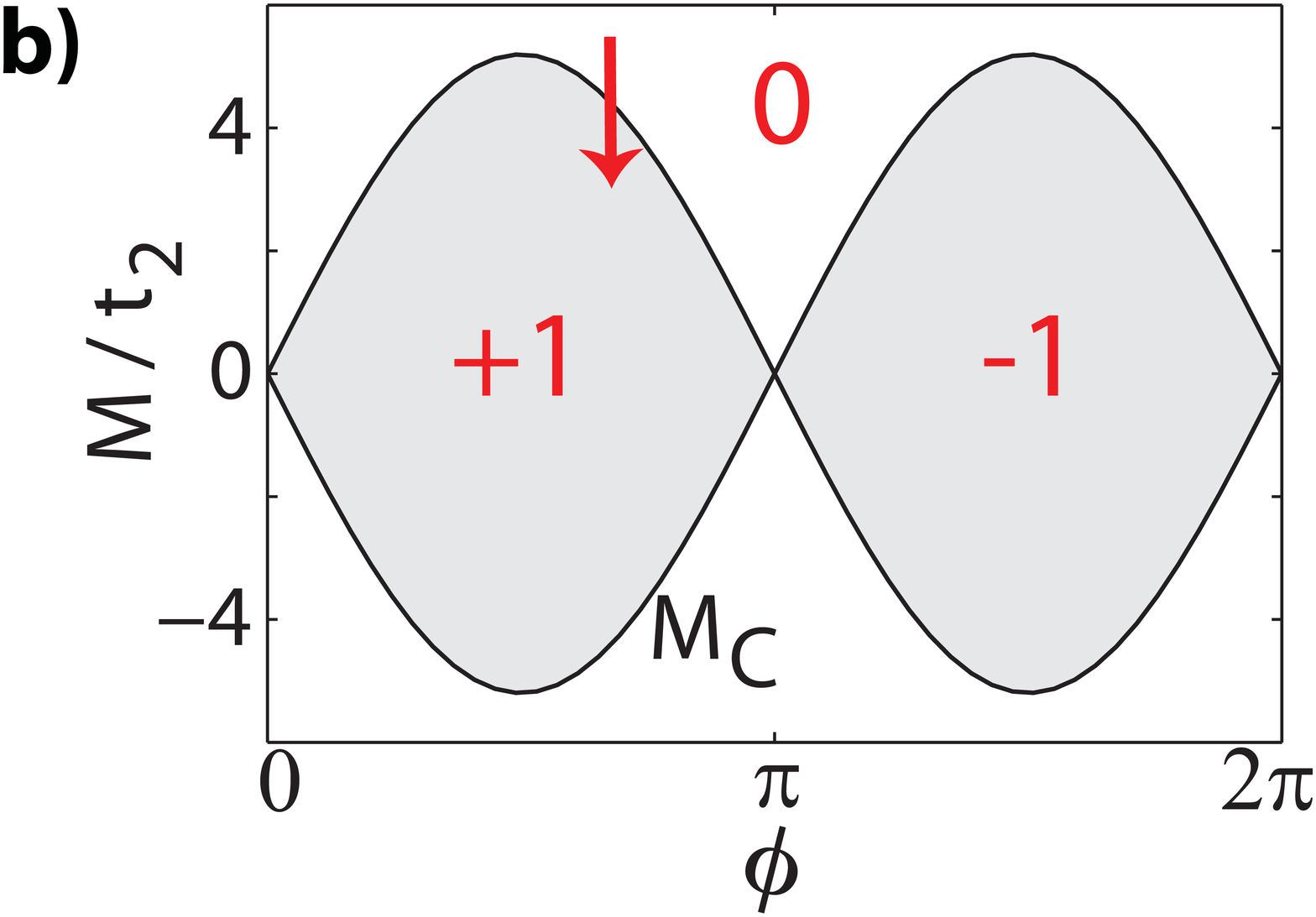} \label{fig Haldane phase diag}}
\caption{ Haldane model. (a) illustrates the matrix elements of the tight-binding Hamiltonian in Eq.(\ref{eq Haldane Hamiltonian}); hoppings $t_1,t_2e^{\pm i\phi}$ and bias $2M$ between $A$ (dark) and $B$ (light) sublattices. (b) Phase diagram showing the Chern number of the lowest band $C=0,\pm1$. The arrow indicates the quench studied in the text. Bands touch at a single Dirac point for $M_C=\pm t_2 3\sqrt{3}\sin\phi$.}
\end{figure}

The Haldane model is a simple yet realistic topological lattice model which realizes a quantum Hall effect in the absence of a net magnetic field \cite{Haldane}. Recent cold-atom realizations \cite{HaldaneExp,Sengstock,MunichHxgnl,NathanNatureRev} provide an ideal platform to study out-of-equilibrium topological phenomena. The static Haldane Hamiltonian describes non-interacting fermions on a honeycomb lattice (see Fig.\ref{fig Haldane lattice})
\begin{align} \label{eq Haldane Hamiltonian}
{\cal H}=-t_1\sum_{<ij>}a_i^{\dag}a_j-t_2\sum_{\ll ij\gg}e^{\pm i\phi}a_i^{\dag}a_j \nonumber\\
+M\sum_A a_i^{\dag}a_i-M\sum_B a_i^{\dag}a_i ,
\end{align}
where $a_i^{\dag}\,(a_i)$ creates (annihilates) a fermion at site $i$. The first term is the nearest-neighbor hopping between the two sublattices, and the second is the next-nearest-neighbor intra-sublattice hopping which breaks time reversal symmetry (TRS). The arrows in Fig.\ref{fig Haldane lattice} represent hopping directions for which the phase gain is positive. An energy offset $M$ of $A$ and $B$ sublattices breaks inversion symmetry (IS). The corresponding phase diagram is given in Fig.\ref{fig Haldane phase diag} where we take $|t_2/t_1|\leq1/3$ to allow the bands to touch but not to overlap \cite{Haldane}. When both $\phi$ and $M$ are zero, the spectrum is graphene-like with two Dirac cones in the Brillouin zone. Breaking TRS opens a gap at the two Dirac points, yielding topological bands with Chern numbers $C=\pm1$. Opening a gap by breaking IS, on the other hand, results in an OI where the Chern numbers vanish. By tuning the competition between the two symmetry breaking terms, one can engineer a topological transition involving only one of the Dirac cones. The critical energy offset of the transition is given by $M_C=\pm t_2 3\sqrt{3}\sin\phi$. The analysis in Section \ref{single dirac section} suggests that a symmetric quench across this boundary will yield a universal non-equilibrium Hall response.

\begin{figure}
\centering
\includegraphics[width=0.47\textwidth]{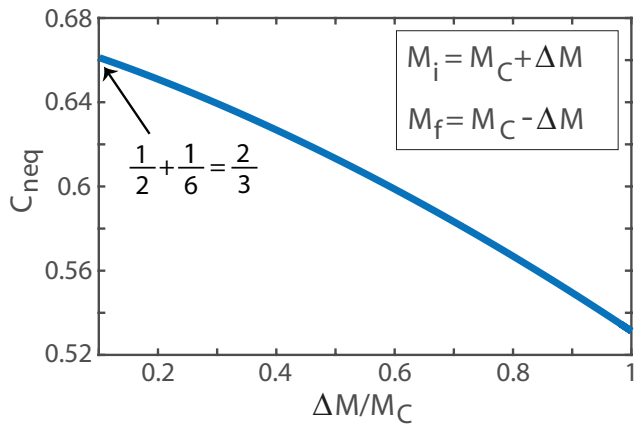}
\caption{ Non-equilibrium Hall response of the Haldane model. The energy offset $M$ is suddenly quenched from $M_i=M_C+\Delta M$ to $M_f=M_C-\Delta M$, where $M_C=t_2 3\sqrt{3}\sin\phi$ is the topological transition point. Here $\phi=0.6\pi$ and $t_2=0.1t_1$. For small quenches one reaches the fractional regime in which $C_{neq}\rightarrow2/3$.  } \label{fig Cneq Haldane}
\end{figure}

We now calculate the Hall response of the Haldane model following a quench that changes the mass sign of one of the Dirac cones in the Brillouin zone. As is simplest in experiments, we quench the energy offset $M$, but a quench of $\phi$ or $t_2$ would also work. We start with a completely filled lower band at energy offset $M_i=M_C+\Delta M$ for $\phi=0.6\pi$ and $t_2=0.1t_1$, where the system is in OI state for $\Delta M>0$. These lattice parameters are representative: we find equivalent results for other parameters. The first Dirac cone initially contributes a Chern number of $-1/2$ where the second one contributes $1/2$, hence the Chern number is zero. We then suddenly change the sublattice bias to $M_f=M_C-\Delta M$, as illustrated in Fig.\ref{fig Haldane phase diag} with the red arrow, and calculate the excitation probabilities at each $k$ in the Brillouin zone. By multiplying the Berry curvatures of the final bands with the occupation probabilities, we calculate the Hall conductivity as in Fig.\ref{fig Cneq Haldane}.

In the limit of $\Delta M\ll M_C$, the transition involves only the first Dirac cone. Since particles near the second Dirac cone are not excited, they continue to contribute $1/2$ to the Chern number. On the other hand, the mass of the first Dirac cone changes sign, so its contribution to the Hall response is $1/6$ as in Sec.\ref{single dirac section}. Together they add up to a Hall response of $2/3$, as seen in Fig.\ref{fig Cneq Haldane}. Note that in this limit, the actual number of particles excited to the upper band is small, but the change in the Hall conductivity is still $2/3$ as the Berry curvature is peaked at the Dirac cone. There are no contributions from the rest of the Brillouin zone as the Berry curvature is negligible and no particles are excited. For larger quenches ($\Delta M\sim M_C$), the Hall conductivity deviates from $2/3$ due to the excitations at the second Dirac cone. For quenches in the other direction, from CI phase to OI phase, the Hall conductivity becomes $C_{neq}=1/2-1/6=1/3$ where the contribution of the first Dirac cone comes with a minus sign.

Experimentally, the local Berry curvature of the bands can be probed by using wave packets \cite{PriceCooper}, which is the measurement of choice in the Zurich experiment \cite{HaldaneExp}. In this method, the wave packet is moved through the Brillouin zone sampling the transverse drift in regions with significant curvature. In quenches on the Haldane model, the action takes place mostly around one of the Dirac cones. If the systems is quenched while the wave packet is passing through the Dirac cone of interest, the resulting transverse drift will reveal the contribution of that Dirac cone alone, namely $1/6$. Since the observation of the universal $1/6$ Hall response does not require a Haldane model, but just the presence of a Dirac cone, it should be also accessible in the hexagonal lattices of the Hamburg \cite{Sengstock} and Munich groups \cite{MunichHxgnl} as long as one can isolate the contribution of a single Dirac cone. In practice, extracting the Chern number from such experiments requires attention to several technical details \cite{PriceCooper,HaldaneExp,BlochHofstadter,Refael, NathanInterfaces}. We discuss an alternative approach in Section \ref{sec harmonic trap}.

\subsection{Beyond Linear Response}  \label{sec linear response}
In an equilibrium state, where the bands are either completely filled or empty, the occupation probability at each point in the Brillouin zone does not change when a force is applied \cite{TKNN}. However, in a non-equilibrium state, the nonuniform distribution will evolve as forces are applied, leading to pronounced nonlinearities in the Hall response. Here we characterize these nonlinearities. Linear response should be valid when the impulse imparted on the atoms $I=E\Delta t$ is small compared to the width of the Dirac cone $\Delta/c$. Here we relax this constraint but still take $E\ll \Delta^2$. In this regime band coherences can be neglected, and one does not encounter the zitterbewegung related effects found in Ref.\cite{Refael} when they considered delta-function impulses.

An infinitesimal impulse $dI\hat{y}$ shifts a particle in the $n^{th}$ band with a momentum $\vec{k}$ to $\vec{k}+dI\hat{y}$, and shifts its center-of-mass in the transverse direction by $dX=\Omega_n(k)dI$. The total displacement of the cloud is found by adding up the contribution from each $(n,k)$ mode; weighted by the probability of occupation $P_n(\vec{k})$. The response to a finite impulse is then calculated by adding up these infinitesimals,
\begin{equation} \label{eq trans displacement}
X(I)=\frac{1}{2\pi}\int_0^I dI^{\prime} \sum_n\int_{BZ} d^2k P_n(\vec{k}-I^{\prime}\hat{y})\Omega_n(\vec{k}).
\end{equation}
The rate of change of this transverse displacement gives the Hall conductivity $\sigma_H=\partial X/\partial I$.

\begin{figure}
\centering
\includegraphics[width=0.47\textwidth]{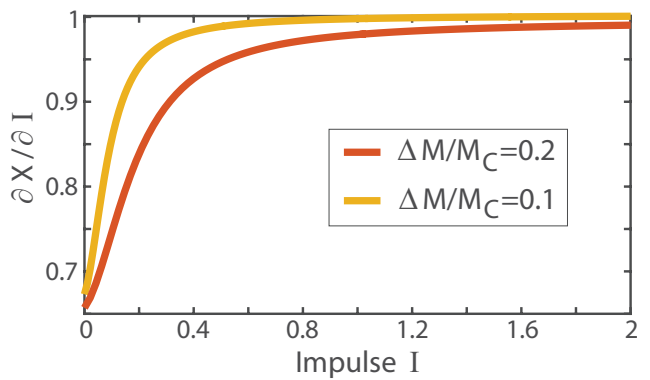}
\caption{ Nonlinear Hall response $C_{neq}=\partial X/\partial I$ following a quench, for transverse displacement $X$ resulting from an impulse $I$. The parameters are adimensionalized by using the unit length of the lattice. The system is quenched between states $\Delta M$ around the transition point $M_C$ as detailed in the text. The Hall response of a non-equilibrium system saturates to equilibrium result ($C=1$ in this case) for large impulse. Linear response regime shrinks for decreasing $\Delta M$. }  \label{fig lin response}
\end{figure}

As in Section \ref{sec Haldane infinite}, we quench the system from the ground state of the OI phase at $M_i=M_C+\Delta M$ to the CI phase at $M_f=M_C-\Delta M$ (where $C=1$) with $t_2=0.1t_1$ and $\phi=0.6\pi$. We numerically calculate the final Berry curvatures $\Omega_n(\vec{k})$ and the probability distributions within the bands $P_n(\vec{k})$. We then apply an impulse $I\hat{y}$ which shifts the probability distributions according to Eq.(\ref{eq trans displacement}). As seen in Fig.\ref{fig lin response}, the resulting Hall conductivity $\sigma_H=\partial X/\partial I$ depends on the applied impulse.

We observe a crossover from a fractional Hall response to an integer response. For large impulses, the excited particles move away from the Dirac point towards $k$-values with negligible curvature. The lower band becomes again completely filled around the Dirac cone and the upper band completely empty. Hence, the Hall response approaches that of the equilibrium system ($C=1$ in this case). In Fig.\ref{fig lin response}, we consider two different values of $\Delta M$. For smaller $\Delta M$, particle excitations and curvature are more confined around the Dirac point, resulting in a smaller linear response regime.

In experiments on equilibrium states, larger impulses are almost always favorable: The bigger the impulse, the larger the transverse drift and hence the larger the signal. In a non-equilibrium setting, however, there is a tension between the desire to have a large signal, and the desire to be in the linear response regime. The strength of the impulse should be chosen carefully.

Up to this point we have been considering an experiment in which the fermionic cloud is initially in its ground state, and hence the Brillouin zone is uniformly filled. One could imagine other scenarios in which this fractional quantized conductance could be explored. For example, one could fill only the states near the Dirac cone (those with $k\lesssim\Delta/c$). In that case, one could observe the $1/6$ Hall response of that Dirac cone. Of course, if the spread of the cloud in $k$-space is too small, or too large, a non-universal value would be found. Many of the Haldane model experiments in Zurich have utilized such partially filled bands \cite{HaldaneExp}.

\subsection{Strip geometry} \label{sec strip}
\begin{figure}
\centering
\includegraphics[width=0.4\textwidth]{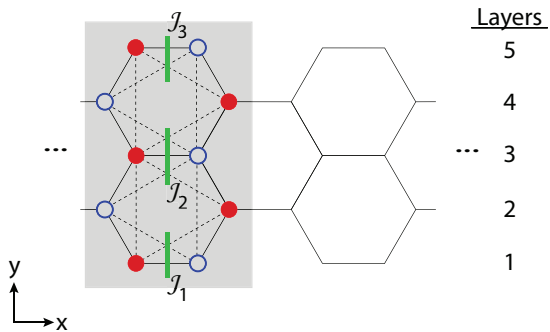}
\caption{ Haldane strip with armchair termination, for parameters introduced in Fig.\ref{fig Haldane lattice}. Unit cell is given by the shaded area. We label the layers in the finite $y$-direction. Green (thick vertical) lines represent how nearest- and next-nearest-neighbor hoppings are assigned into these layers to calculate the current density $\mathcal{J}_{\ell}$. }  \label{fig strip latt}
\end{figure}
The energy spectrum of a finite topological system is fundamentally different than the infinite system due to the presence of protected edge modes. In this section, we investigate how edge states modify both the equilibrium, and non-equilibrium Hall response. We find that for wide enough strips, our results from bulk calculations hold. We also set up notation for exploring the spatial distribution of currents in a harmonic trap. In Sec.\ref{sec harmonic trap}, we show that these currents can be used as a probe of the Hall effect.

We take the system to be infinite in the $x$-direction and finite in the $y$-direction with an armchair termination. We define a `layer' index for each site, which corresponds to its $y$-position as shown in Fig.\ref{fig strip latt}. Under a uniform electric field $E\hat{y}$ in the finite direction, a Hall current $J_H$ flows along the strip in the $x$-direction.

Our Hamiltonian can be expressed as ${\cal H}=-\sum_{ij}t_{i\rightarrow j} a_i^{\dag}a_j+M\sum_A a_i^{\dag}a_i-M\sum_Ba_i^{\dag}a_j$. The matrix element $t_{i\rightarrow j}$ is equal to $t_1$ for neighboring sites, $t_2e^{\pm i\phi}$ for next-neighbor sites, and $0$ for all others. The current from site $i$ to $j$ is then
\begin{equation}
\mathcal{J}_{i\rightarrow j}=2 \text{Im}\{t_{j\rightarrow i} \langle a_i^{\dag}a_j \rangle \}.
\end{equation}
To quantify the spatial distribution of currents, we group the currents according to the layers of $i$ and $j$. For each odd integer $\ell$, we define $\mathcal{J}_{\ell}$ to be the sum of currents either originating from or ending at the $\ell^{th}$ layer, at some fixed $x$-position in the unit cell (see Fig.\ref{fig strip latt}). In an OI state, $J_H=\sum\mathcal{J}_{\ell}$ is always zero even though there may be nonzero currents in individual layers. For a CI phase, however, $J_H$ is finite in the presence of a nonzero electric field. The Hall conductivity of the strip is then $\sigma_H^s=\frac{1}{W} \frac{\partial J_H}{\partial E}$ where $W$ is the width of the strip.

Cold atom experiments can explore strips of varying width, even as small as two layers \cite{BosonicLadder}. Thus, they are well suited to quantifying finite size effects. To analyze these effects in accessing bulk topological properties, here we first consider equilibrium Hall response of a strip in CI regime.

\begin{figure}
\centering
\includegraphics[width=0.47\textwidth]{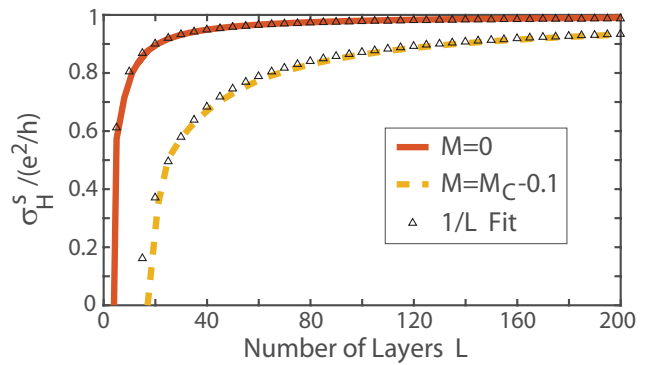}
\caption{ Equilibrium Hall conductivity of the strip as a function of number of layers in the finite direction. The system is in CI state at $M=0$ and $M=M_C-0.1$, for $t_2=0.1t_1$ and $\phi=0.6\pi$. Hall conductivity approaches to the quantized bulk value as $1/L$ for increasing strip width. }  \label{fig power law}
\end{figure}

\begin{figure*}
\centering
\subfigure{\includegraphics[width=0.47\textwidth]{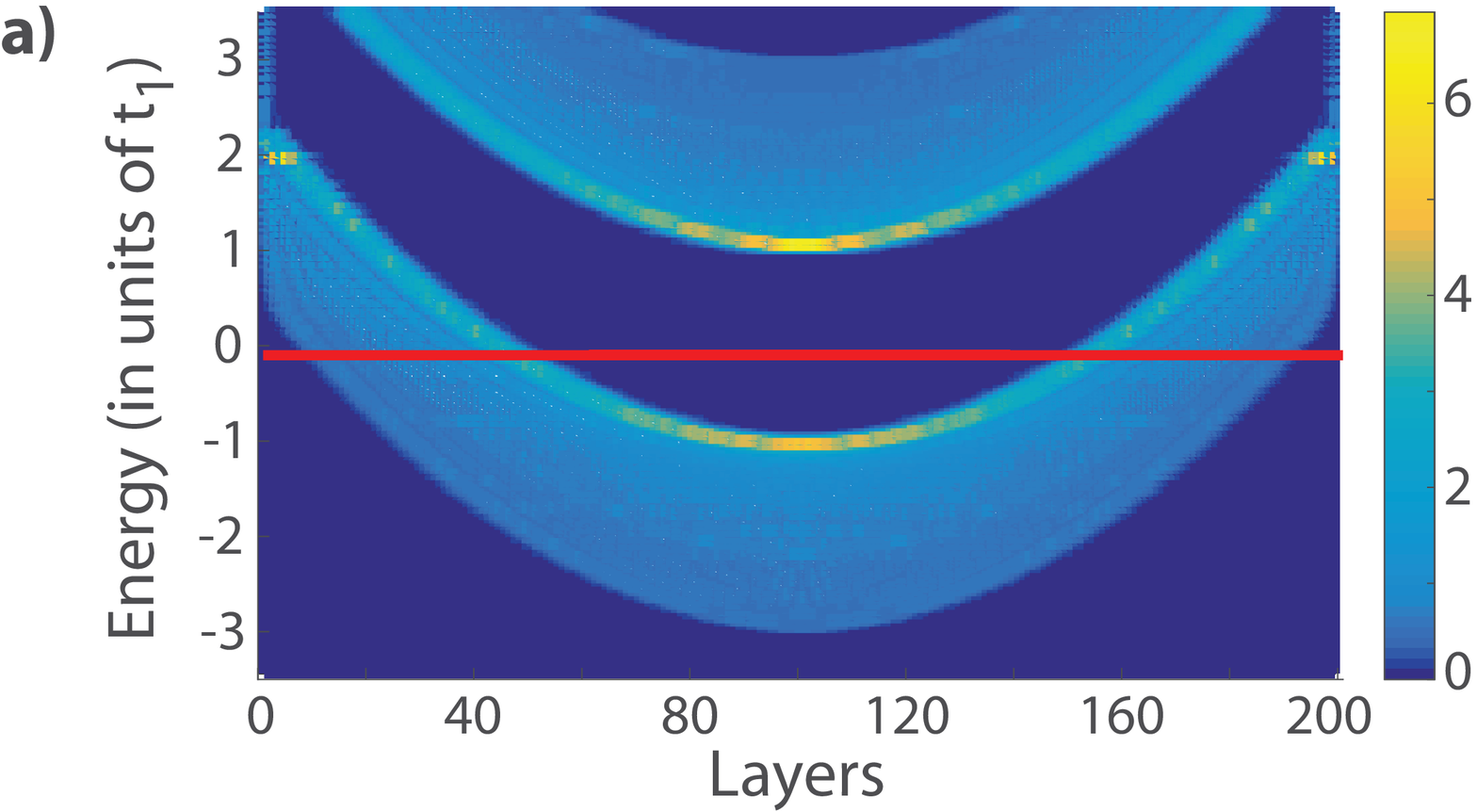} \label{fig LDOS}}
\subfigure{\includegraphics[width=0.47\textwidth]{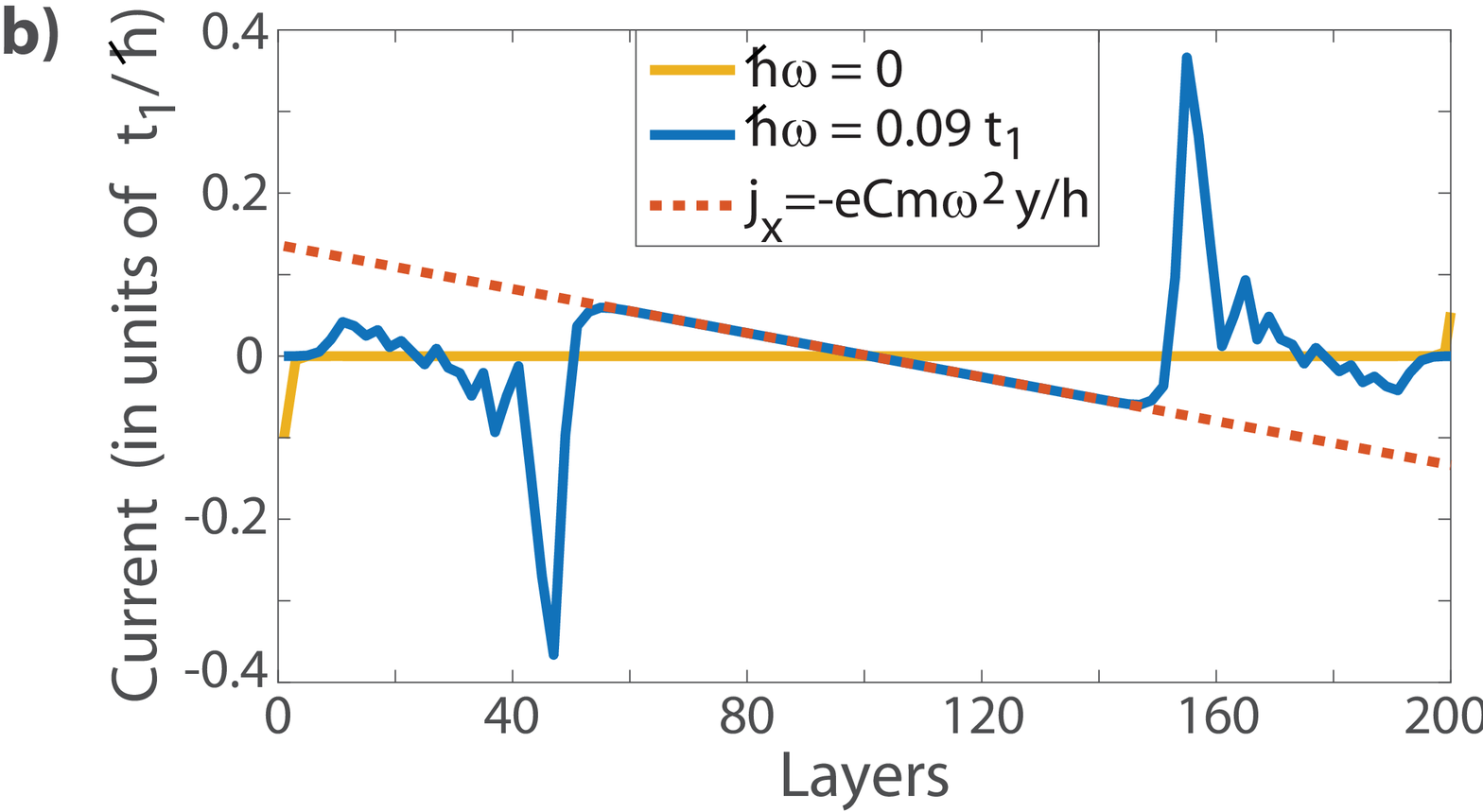} \label{fig trap J}}
\caption{ Effect of a harmonic trap with $\hbar\omega=0.09t_1$ in the finite direction for $M=0,\,t_2=0.2t_1$ and $\phi=0.5\pi$. (a) Local density of states along the finite direction where the center is a CI and the edges are metallic. The Fermi energy is indicated with the red line. (b) Current density flowing along the strip with (dark line) and without (light line) trap, in the absence of an electric field. The force exerted by the trap $F_y=-m\omega^2y$ results in anomalous currents in the central topological region according to $j_x=eCF_y/h$ (dashed line). }  \label{fig trap}
\end{figure*}

We take an $L$ layer strip with $M=0,t_2=0.1t_1$ and $\phi=0.6\pi$. Here, the edge states are well separated from the bulk, and we take the Fermi energy to lie in the bulk gap. An infinite system with these parameters would have a Chern number of $1$. We apply an electric field, and measure the resulting $J_H$. Due to the presence of the hallmark edge states, we find that $\sigma_H$ is somewhat smaller than one would expect from the bulk calculations (see Fig.\ref{fig power law}). The deviation falls off as $1/L$ for large $L$. This is sensible, as the ratio of edge to bulk modes falls off with this same power. We repeat this calculation at $M=M_C-0.1$, where the bulk gap is smaller. Given the larger extent of the edge modes, it is not surprising that we find a larger deviation from bulk behavior.

We now consider a wide strip and study our symmetric quenches of Section \ref{sec Haldane infinite}. We start with a ground state as before in the OI phase at $M_i=M_C+\Delta M$ with the same parameters in the previous section $t_2=0.1t_1$ and $\phi=0.6\pi$. We numerically calculate the initial eigenstates for $L=201$ layers. We then quench the system into the CI phase at $M_f=M_C-\Delta M$ and find the occupation probabilities of the final eigenstates by calculating overlaps with the final eigenstates.

To explore the Hall response, we then add a small electric field $E\hat{y}$, and calculate the currents in our non-equilibrium state. We find that for small quenches $\Delta M/M_C=0.1$ and $0.2$, the resulting Hall response of the strip is $\sigma_H^s=0.66$ and $0.65$ respectively. Hence, in small quenches symmetric around the transition point, the Hall conductivity of a strip successfully reproduces the infinite system result of $2/3$.

\section{Harmonic Trap}  \label{sec harmonic trap}
Nearly all cold atom experiments include a harmonic trap. Naively, this potential complicates the search for topological physics: Even if the center of the cloud is a CI, the edge is always metallic. The metallic edges wash out any potential signals from edge modes, and one might expect them to dominate the Hall response, yielding non-universal, non-quantized results. In this section, we investigate a Haldane model with harmonic confinement added. We find equilibrium currents whose spatial distribution reveals the underlying topological physics. We discuss protocols for detecting these currents, and using them to infer the quantize Hall response.

We take the formalism of Section \ref{sec strip}, and add a potential $H=\sum_i\frac{1}{2}m\omega^2y_i^2$ where $y_i$ is the $y$-position of the $i^{th}$ site. We consider this uni-directional harmonic potential as it makes analyzing the currents simpler. In a cylindrically symmetric harmonic trap, one would observe similar results, with the radial direction playing the role of $y$. In Fig.\ref{fig LDOS}, we show the local density of states of this system, taking $M=0,\,t_2=0.2t_1,\,\phi=0.5\pi$, and $\hbar\omega=0.09t_1$. We take the chemical potential to lie in the bulk gap at the center of the cloud, which will locally be in the CI phase. The metallic edges are characterized by the finite density of states at the Fermi level.

In Fig.\ref{fig trap J}, we show the local equilibrium current $\mathcal{J}_{\ell}$, as a function of position in finite direction in the absence of an electric field. In the insulating region, one sees a current which grows linearly with position in a trap. On the metallic edges, there are large counter-propagating currents. The net current vanishes. The bulk currents are nothing but the quantized anomalous Hall currents coming from the topological band and the trapping potential. The trap is equivalent to a spatially dependent electric field $E_y=-m\omega^2y/e$. Within a local density approximation this leads to a local current density $j_x=\sigma_HE_y=-eCm\omega^2y/h$. Indeed, the slope of the line in Fig.\ref{fig trap J} matches this prediction.

We propose a simple protocol to measure these local currents. One begins with a non-interacting spin polarized Fermi gas in a situation like Fig.\ref{fig LDOS}. Using standard techniques \cite{TakeshiSpinFlip,CornellSpinFlip,KetterleSpinFlip,WrightSpinFlip}, one locally flips the spins of a small number of atoms. Local currents will be apparent in the motions of these spins. Cold atom experiments can explore not only the average current, but also the spatial distribution of the currents \cite{BosonicLadder,Marquardt}. This extra information can disentangle the contributions from the metallic edges and the bulk. Furthermore, if the system is to be quenched, one can still access the Hall response of the topological central region even in the presence of a harmonic trap.

\section{SUMMARY}  \label{sec conclusion}
Out-of-equilibrium systems exhibit new topological phenomena which have no equilibrium counterpart. Here, we report a universal fractional value of the Hall conductivity when the mass sign of a single Dirac cone in a two-band model is suddenly changed. The energy spectrum is independent of the sign of the gap while the eigenstates are sensitive to it. We find that this symmetry results in a contribution to the Hall response of $|C_{neq}|=1/6$ independently from the actual value of the gap. By studying the Haldane model, we illustrate the universality of this result: Any symmetric topological quench dominated by a single Dirac cone will observe this fractional Hall response. We find that the out-of-equilibrium response is highly nonlinear in the impulse. To connect with experiments \cite{HaldaneExp}, we explore finite size effects, and the role of harmonic confinement. By studying finite strips of size $L$, we determine that there are corrections to the quantized conductance which scale as $1/L$. Including the harmonic confinement, we find the equilibrium bulk currents which reveal the quantized Hall effect. We propose an experiment for detecting these currents.

\begin{acknowledgments}
F.N.\"{U}. is supported by The Scientific and Technological Research Council of Turkey (T\"{U}B\.{I}TAK). M.\"{O}.O. would like to thank American Physical Society for International Research Travel Award Program support, and LASSP at Cornell University for their hospitality. E.J.M. acknowledges support from NSFPHY-1508300 and ARO MURI Non-Equilibrium Many-Body Dynamics W9111NF-14-1-0003.
\end{acknowledgments}

\bibliography{HaldaneQuench_arXiv}

\providecommand{\noopsort}[1]{}\providecommand{\singleletter}[1]{#1}%
\begin{thebibliography}{25}%
\makeatletter
\providecommand \@ifxundefined [1]{%
 \@ifx{#1\undefined}
}%
\providecommand \@ifnum [1]{%
 \ifnum #1\expandafter \@firstoftwo
 \else \expandafter \@secondoftwo
 \fi
}%
\providecommand \@ifx [1]{%
 \ifx #1\expandafter \@firstoftwo
 \else \expandafter \@secondoftwo
 \fi
}%
\providecommand \natexlab [1]{#1}%
\providecommand \enquote  [1]{``#1''}%
\providecommand \bibnamefont  [1]{#1}%
\providecommand \bibfnamefont [1]{#1}%
\providecommand \citenamefont [1]{#1}%
\providecommand \href@noop [0]{\@secondoftwo}%
\providecommand \href [0]{\begingroup \@sanitize@url \@href}%
\providecommand \@href[1]{\@@startlink{#1}\@@href}%
\providecommand \@@href[1]{\endgroup#1\@@endlink}%
\providecommand \@sanitize@url [0]{\catcode `\\12\catcode `\$12\catcode
  `\&12\catcode `\#12\catcode `\^12\catcode `\_12\catcode `\%12\relax}%
\providecommand \@@startlink[1]{}%
\providecommand \@@endlink[0]{}%
\providecommand \url  [0]{\begingroup\@sanitize@url \@url }%
\providecommand \@url [1]{\endgroup\@href {#1}{\urlprefix }}%
\providecommand \urlprefix  [0]{URL }%
\providecommand \Eprint [0]{\href }%
\providecommand \doibase [0]{http://dx.doi.org/}%
\providecommand \selectlanguage [0]{\@gobble}%
\providecommand \bibinfo  [0]{\@secondoftwo}%
\providecommand \bibfield  [0]{\@secondoftwo}%
\providecommand \translation [1]{[#1]}%
\providecommand \BibitemOpen [0]{}%
\providecommand \bibitemStop [0]{}%
\providecommand \bibitemNoStop [0]{.\EOS\space}%
\providecommand \EOS [0]{\spacefactor3000\relax}%
\providecommand \BibitemShut  [1]{\csname bibitem#1\endcsname}%
\let\auto@bib@innerbib\@empty
\bibitem [{\citenamefont {D'Alessio}\ and\ \citenamefont
  {Rigol}(2014)}]{Rigol}%
  \BibitemOpen
  \bibfield  {author} {\bibinfo {author} {\bibfnamefont {Luca}\ \bibnamefont
  {D'Alessio}}\ and\ \bibinfo {author} {\bibfnamefont {Marcos}\ \bibnamefont
  {Rigol}},\ }\bibfield  {title} {\enquote {\bibinfo {title} {Dynamical
  preparation of floquet chern insulators},}\ }\href {\doibase
  10.1038/ncomms9336} {\bibfield  {journal} {\bibinfo  {journal} {Nature
  Communications}\ }\textbf {\bibinfo {volume} {6}},\ \bibinfo {pages} {8336}
  (\bibinfo {year} {2014})}\BibitemShut {NoStop}%
\bibitem [{\citenamefont {Caio}\ \emph {et~al.}(2015)\citenamefont {Caio},
  \citenamefont {Cooper},\ and\ \citenamefont {Bhaseen}}]{CooperPRL}%
  \BibitemOpen
  \bibfield  {author} {\bibinfo {author} {\bibfnamefont {M.~D.}\ \bibnamefont
  {Caio}}, \bibinfo {author} {\bibfnamefont {N.~R.}\ \bibnamefont {Cooper}}, \
  and\ \bibinfo {author} {\bibfnamefont {M.~J.}\ \bibnamefont {Bhaseen}},\
  }\bibfield  {title} {\enquote {\bibinfo {title} {Quantum quenches in chern
  insulators},}\ }\href {\doibase 10.1103/PhysRevLett.115.236403} {\bibfield
  {journal} {\bibinfo  {journal} {Phys. Rev. Lett.}\ }\textbf {\bibinfo
  {volume} {115}},\ \bibinfo {pages} {236403} (\bibinfo {year}
  {2015})}\BibitemShut {NoStop}%
\bibitem [{\citenamefont {Wang}\ \emph {et~al.}(2016)\citenamefont {Wang},
  \citenamefont {Schmitt},\ and\ \citenamefont {Kehrein}}]{KehreinPRB}%
  \BibitemOpen
  \bibfield  {author} {\bibinfo {author} {\bibfnamefont {Pei}\ \bibnamefont
  {Wang}}, \bibinfo {author} {\bibfnamefont {Markus}\ \bibnamefont {Schmitt}},
  \ and\ \bibinfo {author} {\bibfnamefont {Stefan}\ \bibnamefont {Kehrein}},\
  }\bibfield  {title} {\enquote {\bibinfo {title} {Universal nonanalytic
  behavior of the hall conductance in a chern insulator at the topologically
  driven nonequilibrium phase transition},}\ }\href {\doibase
  10.1103/PhysRevB.93.085134} {\bibfield  {journal} {\bibinfo  {journal} {Phys.
  Rev. B}\ }\textbf {\bibinfo {volume} {93}},\ \bibinfo {pages} {085134}
  (\bibinfo {year} {2016})}\BibitemShut {NoStop}%
\bibitem [{\citenamefont {Fl\"{a}schner}\ \emph {et~al.}(2016)\citenamefont
  {Fl\"{a}schner}, \citenamefont {Vogel}, \citenamefont {Tarnowski},
  \citenamefont {Rem}, \citenamefont {Lühmann}, \citenamefont {Heyl},
  \citenamefont {Budich}, \citenamefont {Mathey}, \citenamefont {Sengstock},\
  and\ \citenamefont {Weitenberg}}]{Sengstock}%
  \BibitemOpen
  \bibfield  {author} {\bibinfo {author} {\bibfnamefont {Nick}\ \bibnamefont
  {Fl\"{a}schner}}, \bibinfo {author} {\bibfnamefont {Dominik}\ \bibnamefont
  {Vogel}}, \bibinfo {author} {\bibfnamefont {Matthias}\ \bibnamefont
  {Tarnowski}}, \bibinfo {author} {\bibfnamefont {Benno~S.}\ \bibnamefont
  {Rem}}, \bibinfo {author} {\bibfnamefont {Dirk-S\"{o}ren}\ \bibnamefont
  {Lühmann}}, \bibinfo {author} {\bibfnamefont {Markus}\ \bibnamefont {Heyl}},
  \bibinfo {author} {\bibfnamefont {Jan~Carl}\ \bibnamefont {Budich}}, \bibinfo
  {author} {\bibfnamefont {Ludwig}\ \bibnamefont {Mathey}}, \bibinfo {author}
  {\bibfnamefont {Klaus}\ \bibnamefont {Sengstock}}, \ and\ \bibinfo {author}
  {\bibfnamefont {Christof}\ \bibnamefont {Weitenberg}},\ }\bibfield  {title}
  {\enquote {\bibinfo {title} {Observation of a dynamical topological phase
  transition},}\ }\href {http://arxiv.org/abs/1608.05616} {\bibfield  {journal}
  {\bibinfo  {journal} {arXiv:1608.05616}\ } (\bibinfo {year}
  {2016})}\BibitemShut {NoStop}%
\bibitem [{\citenamefont {Wilson}\ \emph {et~al.}(2016)\citenamefont {Wilson},
  \citenamefont {Song},\ and\ \citenamefont {Refael}}]{Refael}%
  \BibitemOpen
  \bibfield  {author} {\bibinfo {author} {\bibfnamefont {Justin~H.}\
  \bibnamefont {Wilson}}, \bibinfo {author} {\bibfnamefont {Justin C.~W.}\
  \bibnamefont {Song}}, \ and\ \bibinfo {author} {\bibfnamefont {Gil}\
  \bibnamefont {Refael}},\ }\bibfield  {title} {\enquote {\bibinfo {title}
  {Persistent hall response in a quantum quench},}\ }\href
  {http://arxiv.org/abs/1603.01621} {\bibfield  {journal} {\bibinfo  {journal}
  {arXiv:1603.01621}\ } (\bibinfo {year} {2016})}\BibitemShut {NoStop}%
\bibitem [{\citenamefont {Hu}\ \emph {et~al.}(2016)\citenamefont {Hu},
  \citenamefont {Zoller},\ and\ \citenamefont {Budich}}]{DynBuildup}%
  \BibitemOpen
  \bibfield  {author} {\bibinfo {author} {\bibfnamefont {Ying}\ \bibnamefont
  {Hu}}, \bibinfo {author} {\bibfnamefont {Peter}\ \bibnamefont {Zoller}}, \
  and\ \bibinfo {author} {\bibfnamefont {Jan~Carl}\ \bibnamefont {Budich}},\
  }\bibfield  {title} {\enquote {\bibinfo {title} {Dynamical buildup of a
  quantized hall response from nontopological states},}\ }\href {\doibase
  10.1103/PhysRevLett.117.126803} {\bibfield  {journal} {\bibinfo  {journal}
  {Phys. Rev. Lett.}\ }\textbf {\bibinfo {volume} {117}},\ \bibinfo {pages}
  {126803} (\bibinfo {year} {2016})}\BibitemShut {NoStop}%
\bibitem [{\citenamefont {Caio}\ \emph {et~al.}(2016)\citenamefont {Caio},
  \citenamefont {Cooper},\ and\ \citenamefont {Bhaseen}}]{Cooper2}%
  \BibitemOpen
  \bibfield  {author} {\bibinfo {author} {\bibfnamefont {M.~D.}\ \bibnamefont
  {Caio}}, \bibinfo {author} {\bibfnamefont {N.~R.}\ \bibnamefont {Cooper}}, \
  and\ \bibinfo {author} {\bibfnamefont {M.~J.}\ \bibnamefont {Bhaseen}},\
  }\bibfield  {title} {\enquote {\bibinfo {title} {Hall response and edge
  current dynamics in chern insulators out of equilibrium},}\ }\href {\doibase
  10.1103/PhysRevB.94.155104} {\bibfield  {journal} {\bibinfo  {journal} {Phys.
  Rev. B}\ }\textbf {\bibinfo {volume} {94}},\ \bibinfo {pages} {155104}
  (\bibinfo {year} {2016})}\BibitemShut {NoStop}%
\bibitem [{\citenamefont {Foster}\ \emph {et~al.}(2014)\citenamefont {Foster},
  \citenamefont {Gurarie}, \citenamefont {Dzero},\ and\ \citenamefont
  {Yuzbashyan}}]{Yuzbashyan}%
  \BibitemOpen
  \bibfield  {author} {\bibinfo {author} {\bibfnamefont {Matthew~S.}\
  \bibnamefont {Foster}}, \bibinfo {author} {\bibfnamefont {Victor}\
  \bibnamefont {Gurarie}}, \bibinfo {author} {\bibfnamefont {Maxim}\
  \bibnamefont {Dzero}}, \ and\ \bibinfo {author} {\bibfnamefont {Emil~A.}\
  \bibnamefont {Yuzbashyan}},\ }\bibfield  {title} {\enquote {\bibinfo {title}
  {Quench-induced floquet topological $p$-wave superfluids},}\ }\href {\doibase
  10.1103/PhysRevLett.113.076403} {\bibfield  {journal} {\bibinfo  {journal}
  {Phys. Rev. Lett.}\ }\textbf {\bibinfo {volume} {113}},\ \bibinfo {pages}
  {076403} (\bibinfo {year} {2014})}\BibitemShut {NoStop}%
\bibitem [{\citenamefont {Dehghani}\ \emph {et~al.}(2015)\citenamefont
  {Dehghani}, \citenamefont {Oka},\ and\ \citenamefont {Mitra}}]{AditiPRB}%
  \BibitemOpen
  \bibfield  {author} {\bibinfo {author} {\bibfnamefont {Hossein}\ \bibnamefont
  {Dehghani}}, \bibinfo {author} {\bibfnamefont {Takashi}\ \bibnamefont {Oka}},
  \ and\ \bibinfo {author} {\bibfnamefont {Aditi}\ \bibnamefont {Mitra}},\
  }\bibfield  {title} {\enquote {\bibinfo {title} {Out-of-equilibrium electrons
  and the hall conductance of a floquet topological insulator},}\ }\href
  {\doibase 10.1103/PhysRevB.91.155422} {\bibfield  {journal} {\bibinfo
  {journal} {Phys. Rev. B}\ }\textbf {\bibinfo {volume} {91}},\ \bibinfo
  {pages} {155422} (\bibinfo {year} {2015})}\BibitemShut {NoStop}%
\bibitem [{\citenamefont {Jotzu}\ \emph {et~al.}(2014)\citenamefont {Jotzu},
  \citenamefont {Messer}, \citenamefont {Desbuquois}, \citenamefont {Lebrat},
  \citenamefont {Uehlinger}, \citenamefont {Greif},\ and\ \citenamefont
  {Esslinger}}]{HaldaneExp}%
  \BibitemOpen
  \bibfield  {author} {\bibinfo {author} {\bibfnamefont {Gregor}\ \bibnamefont
  {Jotzu}}, \bibinfo {author} {\bibfnamefont {Michael}\ \bibnamefont {Messer}},
  \bibinfo {author} {\bibfnamefont {R\'{e}mi}\ \bibnamefont {Desbuquois}},
  \bibinfo {author} {\bibfnamefont {Martin}\ \bibnamefont {Lebrat}}, \bibinfo
  {author} {\bibfnamefont {Thomas}\ \bibnamefont {Uehlinger}}, \bibinfo
  {author} {\bibfnamefont {Daniel}\ \bibnamefont {Greif}}, \ and\ \bibinfo
  {author} {\bibfnamefont {Tilman}\ \bibnamefont {Esslinger}},\ }\bibfield
  {title} {\enquote {\bibinfo {title} {Experimental realization of the
  topological haldane model with ultracold fermions},}\ }\href {\doibase
  10.1038/nature13915} {\bibfield  {journal} {\bibinfo  {journal} {Nature}\
  }\textbf {\bibinfo {volume} {515}},\ \bibinfo {pages} {237} (\bibinfo {year}
  {2014})}\BibitemShut {NoStop}%
\bibitem [{\citenamefont {Haldane}(1988)}]{Haldane}%
  \BibitemOpen
  \bibfield  {author} {\bibinfo {author} {\bibfnamefont {F.~D.~M.}\
  \bibnamefont {Haldane}},\ }\bibfield  {title} {\enquote {\bibinfo {title}
  {Model for a quantum hall effect without landau levels: Condensed-matter
  realization of the "parity anomaly"},}\ }\href {\doibase
  10.1103/PhysRevLett.61.2015} {\bibfield  {journal} {\bibinfo  {journal}
  {Phys. Rev. Lett.}\ }\textbf {\bibinfo {volume} {61}},\ \bibinfo {pages}
  {2015--2018} (\bibinfo {year} {1988})}\BibitemShut {NoStop}%
\bibitem [{\citenamefont {Xiao}\ \emph {et~al.}(2010)\citenamefont {Xiao},
  \citenamefont {Chang},\ and\ \citenamefont {Niu}}]{XiaoRev}%
  \BibitemOpen
  \bibfield  {author} {\bibinfo {author} {\bibfnamefont {Di}~\bibnamefont
  {Xiao}}, \bibinfo {author} {\bibfnamefont {Ming-Che}\ \bibnamefont {Chang}},
  \ and\ \bibinfo {author} {\bibfnamefont {Qian}\ \bibnamefont {Niu}},\
  }\bibfield  {title} {\enquote {\bibinfo {title} {Berry phase effects on
  electronic properties},}\ }\href {\doibase 10.1103/RevModPhys.82.1959}
  {\bibfield  {journal} {\bibinfo  {journal} {Rev. Mod. Phys.}\ }\textbf
  {\bibinfo {volume} {82}},\ \bibinfo {pages} {1959--2007} (\bibinfo {year}
  {2010})}\BibitemShut {NoStop}%
\bibitem [{\citenamefont {Hasan}\ and\ \citenamefont
  {Kane}(2010)}]{HasanKaneRev}%
  \BibitemOpen
  \bibfield  {author} {\bibinfo {author} {\bibfnamefont {M.~Z.}\ \bibnamefont
  {Hasan}}\ and\ \bibinfo {author} {\bibfnamefont {C.~L.}\ \bibnamefont
  {Kane}},\ }\bibfield  {title} {\enquote {\bibinfo {title} {Colloquium:
  Topological insulators},}\ }\href {\doibase 10.1103/RevModPhys.82.3045}
  {\bibfield  {journal} {\bibinfo  {journal} {Rev. Mod. Phys.}\ }\textbf
  {\bibinfo {volume} {82}},\ \bibinfo {pages} {3045--3067} (\bibinfo {year}
  {2010})}\BibitemShut {NoStop}%
\bibitem [{\citenamefont {Thouless}\ \emph {et~al.}(1982)\citenamefont
  {Thouless}, \citenamefont {Kohmoto}, \citenamefont {Nightingale},\ and\
  \citenamefont {den Nijs}}]{TKNN}%
  \BibitemOpen
  \bibfield  {author} {\bibinfo {author} {\bibfnamefont {D.~J.}\ \bibnamefont
  {Thouless}}, \bibinfo {author} {\bibfnamefont {M.}~\bibnamefont {Kohmoto}},
  \bibinfo {author} {\bibfnamefont {M.~P.}\ \bibnamefont {Nightingale}}, \ and\
  \bibinfo {author} {\bibfnamefont {M.}~\bibnamefont {den Nijs}},\ }\bibfield
  {title} {\enquote {\bibinfo {title} {Quantized hall conductance in a
  two-dimensional periodic potential},}\ }\href {\doibase
  10.1103/PhysRevLett.49.405} {\bibfield  {journal} {\bibinfo  {journal} {Phys.
  Rev. Lett.}\ }\textbf {\bibinfo {volume} {49}},\ \bibinfo {pages} {405--408}
  (\bibinfo {year} {1982})}\BibitemShut {NoStop}%
\bibitem [{\citenamefont {Duca}\ \emph {et~al.}(2015)\citenamefont {Duca},
  \citenamefont {Li}, \citenamefont {Reitter}, \citenamefont {Bloch},
  \citenamefont {Schleier-Smith},\ and\ \citenamefont
  {Schneider}}]{MunichHxgnl}%
  \BibitemOpen
  \bibfield  {author} {\bibinfo {author} {\bibfnamefont {L.}~\bibnamefont
  {Duca}}, \bibinfo {author} {\bibfnamefont {T.}~\bibnamefont {Li}}, \bibinfo
  {author} {\bibfnamefont {M.}~\bibnamefont {Reitter}}, \bibinfo {author}
  {\bibfnamefont {I.}~\bibnamefont {Bloch}}, \bibinfo {author} {\bibfnamefont
  {M.}~\bibnamefont {Schleier-Smith}}, \ and\ \bibinfo {author} {\bibfnamefont
  {U.}~\bibnamefont {Schneider}},\ }\bibfield  {title} {\enquote {\bibinfo
  {title} {An aharonov-bohm interferometer for determining bloch band
  topology},}\ }\href {\doibase 10.1126/science.1259052} {\bibfield  {journal}
  {\bibinfo  {journal} {Science}\ }\textbf {\bibinfo {volume} {347}},\ \bibinfo
  {pages} {288--292} (\bibinfo {year} {2015})}\BibitemShut {NoStop}%
\bibitem [{\citenamefont {Goldman}\ \emph
  {et~al.}(2016{\natexlab{a}})\citenamefont {Goldman}, \citenamefont {Budich},\
  and\ \citenamefont {Zoller}}]{NathanNatureRev}%
  \BibitemOpen
  \bibfield  {author} {\bibinfo {author} {\bibfnamefont {N.}~\bibnamefont
  {Goldman}}, \bibinfo {author} {\bibfnamefont {J.~C.}\ \bibnamefont {Budich}},
  \ and\ \bibinfo {author} {\bibfnamefont {P.}~\bibnamefont {Zoller}},\
  }\bibfield  {title} {\enquote {\bibinfo {title} {Topological quantum matter
  with ultracold gases in optical lattices},}\ }\href {\doibase
  10.1038/nphys3803} {\bibfield  {journal} {\bibinfo  {journal} {Nature
  Physics}\ }\textbf {\bibinfo {volume} {12}},\ \bibinfo {pages} {639}
  (\bibinfo {year} {2016}{\natexlab{a}})}\BibitemShut {NoStop}%
\bibitem [{\citenamefont {Price}\ and\ \citenamefont
  {Cooper}(2012)}]{PriceCooper}%
  \BibitemOpen
  \bibfield  {author} {\bibinfo {author} {\bibfnamefont {H.~M.}\ \bibnamefont
  {Price}}\ and\ \bibinfo {author} {\bibfnamefont {N.~R.}\ \bibnamefont
  {Cooper}},\ }\bibfield  {title} {\enquote {\bibinfo {title} {Mapping the
  berry curvature from semiclassical dynamics in optical lattices},}\ }\href
  {\doibase 10.1103/PhysRevA.85.033620} {\bibfield  {journal} {\bibinfo
  {journal} {Phys. Rev. A}\ }\textbf {\bibinfo {volume} {85}},\ \bibinfo
  {pages} {033620} (\bibinfo {year} {2012})}\BibitemShut {NoStop}%
\bibitem [{\citenamefont {Aidelsburger}\ \emph {et~al.}(2015)\citenamefont
  {Aidelsburger}, \citenamefont {Lohse}, \citenamefont {Schweizer},
  \citenamefont {Atala}, \citenamefont {Barreiro}, \citenamefont
  {Nascimb\`{e}ne}, \citenamefont {Cooper}, \citenamefont {Bloch},\ and\
  \citenamefont {Goldman}}]{BlochHofstadter}%
  \BibitemOpen
  \bibfield  {author} {\bibinfo {author} {\bibfnamefont {M.}~\bibnamefont
  {Aidelsburger}}, \bibinfo {author} {\bibfnamefont {M.}~\bibnamefont {Lohse}},
  \bibinfo {author} {\bibfnamefont {C.}~\bibnamefont {Schweizer}}, \bibinfo
  {author} {\bibfnamefont {M.}~\bibnamefont {Atala}}, \bibinfo {author}
  {\bibfnamefont {J.~T.}\ \bibnamefont {Barreiro}}, \bibinfo {author}
  {\bibfnamefont {S.}~\bibnamefont {Nascimb\`{e}ne}}, \bibinfo {author}
  {\bibfnamefont {N.~R.}\ \bibnamefont {Cooper}}, \bibinfo {author}
  {\bibfnamefont {I.}~\bibnamefont {Bloch}}, \ and\ \bibinfo {author}
  {\bibfnamefont {N.}~\bibnamefont {Goldman}},\ }\bibfield  {title} {\enquote
  {\bibinfo {title} {Measuring the chern number of hofstadter bands with
  ultracold bosonic atoms},}\ }\href {\doibase 10.1038/nphys3171} {\bibfield
  {journal} {\bibinfo  {journal} {Nature Physics}\ }\textbf {\bibinfo {volume}
  {11}},\ \bibinfo {pages} {162} (\bibinfo {year} {2015})}\BibitemShut
  {NoStop}%
\bibitem [{\citenamefont {Goldman}\ \emph
  {et~al.}(2016{\natexlab{b}})\citenamefont {Goldman}, \citenamefont {Jotzu},
  \citenamefont {Messer}, \citenamefont {Görg}, \citenamefont {Desbuquois},\
  and\ \citenamefont {Esslinger}}]{NathanInterfaces}%
  \BibitemOpen
  \bibfield  {author} {\bibinfo {author} {\bibfnamefont {N.}~\bibnamefont
  {Goldman}}, \bibinfo {author} {\bibfnamefont {G.}~\bibnamefont {Jotzu}},
  \bibinfo {author} {\bibfnamefont {M.}~\bibnamefont {Messer}}, \bibinfo
  {author} {\bibfnamefont {F.}~\bibnamefont {Görg}}, \bibinfo {author}
  {\bibfnamefont {R.}~\bibnamefont {Desbuquois}}, \ and\ \bibinfo {author}
  {\bibfnamefont {T.}~\bibnamefont {Esslinger}},\ }\bibfield  {title} {\enquote
  {\bibinfo {title} {Creating topological interfaces and detecting chiral edge
  modes in a 2d optical lattice},}\ }\href {http://arxiv.org/abs/1606.00015}
  {\bibfield  {journal} {\bibinfo  {journal} {arXiv:1606.00015}\ } (\bibinfo
  {year} {2016}{\natexlab{b}})}\BibitemShut {NoStop}%
\bibitem [{\citenamefont {Atala}\ \emph {et~al.}(2014)\citenamefont {Atala},
  \citenamefont {Aidelsburger}, \citenamefont {Lohse}, \citenamefont
  {Barreiro}, \citenamefont {Paredes},\ and\ \citenamefont
  {Bloch}}]{BosonicLadder}%
  \BibitemOpen
  \bibfield  {author} {\bibinfo {author} {\bibfnamefont {Marcos}\ \bibnamefont
  {Atala}}, \bibinfo {author} {\bibfnamefont {Monika}\ \bibnamefont
  {Aidelsburger}}, \bibinfo {author} {\bibfnamefont {Michael}\ \bibnamefont
  {Lohse}}, \bibinfo {author} {\bibfnamefont {Julio~T.}\ \bibnamefont
  {Barreiro}}, \bibinfo {author} {\bibfnamefont {Bel\'{e}n}\ \bibnamefont
  {Paredes}}, \ and\ \bibinfo {author} {\bibfnamefont {Immanuel}\ \bibnamefont
  {Bloch}},\ }\bibfield  {title} {\enquote {\bibinfo {title} {Observation of
  chiral currents with ultracold atoms in bosonic ladders},}\ }\href {\doibase
  10.1038/nphys2998} {\bibfield  {journal} {\bibinfo  {journal} {Nature
  Physics}\ }\textbf {\bibinfo {volume} {10}},\ \bibinfo {pages} {588--593}
  (\bibinfo {year} {2014})}\BibitemShut {NoStop}%
\bibitem [{\citenamefont {Fukuhara}\ \emph {et~al.}(2013)\citenamefont
  {Fukuhara}, \citenamefont {Kantian}, \citenamefont {Endres}, \citenamefont
  {Cheneau}, \citenamefont {Schauß}, \citenamefont {Hild}, \citenamefont
  {Bellem}, \citenamefont {Schollwöck}, \citenamefont {Giamarchi},
  \citenamefont {Gross}, \citenamefont {Bloch},\ and\ \citenamefont
  {Kuhr}}]{TakeshiSpinFlip}%
  \BibitemOpen
  \bibfield  {author} {\bibinfo {author} {\bibfnamefont {Takeshi}\ \bibnamefont
  {Fukuhara}}, \bibinfo {author} {\bibfnamefont {Adrian}\ \bibnamefont
  {Kantian}}, \bibinfo {author} {\bibfnamefont {Manuel}\ \bibnamefont
  {Endres}}, \bibinfo {author} {\bibfnamefont {Marc}\ \bibnamefont {Cheneau}},
  \bibinfo {author} {\bibfnamefont {Peter}\ \bibnamefont {Schauß}}, \bibinfo
  {author} {\bibfnamefont {Sebastian}\ \bibnamefont {Hild}}, \bibinfo {author}
  {\bibfnamefont {David}\ \bibnamefont {Bellem}}, \bibinfo {author}
  {\bibfnamefont {Ulrich}\ \bibnamefont {Schollwöck}}, \bibinfo {author}
  {\bibfnamefont {Thierry}\ \bibnamefont {Giamarchi}}, \bibinfo {author}
  {\bibfnamefont {Christian}\ \bibnamefont {Gross}}, \bibinfo {author}
  {\bibfnamefont {Immanuel}\ \bibnamefont {Bloch}}, \ and\ \bibinfo {author}
  {\bibfnamefont {Stefan}\ \bibnamefont {Kuhr}},\ }\bibfield  {title} {\enquote
  {\bibinfo {title} {Quantum dynamics of a mobile spin impurity},}\ }\href
  {\doibase 10.1038/nphys2561} {\bibfield  {journal} {\bibinfo  {journal}
  {Nature Physics}\ }\textbf {\bibinfo {volume} {9}},\ \bibinfo {pages}
  {235--241} (\bibinfo {year} {2013})}\BibitemShut {NoStop}%
\bibitem [{\citenamefont {Matthews}\ \emph {et~al.}(1999)\citenamefont
  {Matthews}, \citenamefont {Anderson}, \citenamefont {Haljan}, \citenamefont
  {Hall}, \citenamefont {Wieman},\ and\ \citenamefont
  {Cornell}}]{CornellSpinFlip}%
  \BibitemOpen
  \bibfield  {author} {\bibinfo {author} {\bibfnamefont {M.~R.}\ \bibnamefont
  {Matthews}}, \bibinfo {author} {\bibfnamefont {B.~P.}\ \bibnamefont
  {Anderson}}, \bibinfo {author} {\bibfnamefont {P.~C.}\ \bibnamefont
  {Haljan}}, \bibinfo {author} {\bibfnamefont {D.~S.}\ \bibnamefont {Hall}},
  \bibinfo {author} {\bibfnamefont {C.~E.}\ \bibnamefont {Wieman}}, \ and\
  \bibinfo {author} {\bibfnamefont {E.~A.}\ \bibnamefont {Cornell}},\
  }\bibfield  {title} {\enquote {\bibinfo {title} {Vortices in a bose-einstein
  condensate},}\ }\href {\doibase 10.1103/PhysRevLett.83.2498} {\bibfield
  {journal} {\bibinfo  {journal} {Phys. Rev. Lett.}\ }\textbf {\bibinfo
  {volume} {83}},\ \bibinfo {pages} {2498--2501} (\bibinfo {year}
  {1999})}\BibitemShut {NoStop}%
\bibitem [{\citenamefont {Chikkatur}\ \emph {et~al.}(2000)\citenamefont
  {Chikkatur}, \citenamefont {G\"orlitz}, \citenamefont {Stamper-Kurn},
  \citenamefont {Inouye}, \citenamefont {Gupta},\ and\ \citenamefont
  {Ketterle}}]{KetterleSpinFlip}%
  \BibitemOpen
  \bibfield  {author} {\bibinfo {author} {\bibfnamefont {A.~P.}\ \bibnamefont
  {Chikkatur}}, \bibinfo {author} {\bibfnamefont {A.}~\bibnamefont
  {G\"orlitz}}, \bibinfo {author} {\bibfnamefont {D.~M.}\ \bibnamefont
  {Stamper-Kurn}}, \bibinfo {author} {\bibfnamefont {S.}~\bibnamefont
  {Inouye}}, \bibinfo {author} {\bibfnamefont {S.}~\bibnamefont {Gupta}}, \
  and\ \bibinfo {author} {\bibfnamefont {W.}~\bibnamefont {Ketterle}},\
  }\bibfield  {title} {\enquote {\bibinfo {title} {Suppression and enhancement
  of impurity scattering in a bose-einstein condensate},}\ }\href {\doibase
  10.1103/PhysRevLett.85.483} {\bibfield  {journal} {\bibinfo  {journal} {Phys.
  Rev. Lett.}\ }\textbf {\bibinfo {volume} {85}},\ \bibinfo {pages} {483--486}
  (\bibinfo {year} {2000})}\BibitemShut {NoStop}%
\bibitem [{\citenamefont {Wright}\ \emph {et~al.}(2008)\citenamefont {Wright},
  \citenamefont {Leslie},\ and\ \citenamefont {Bigelow}}]{WrightSpinFlip}%
  \BibitemOpen
  \bibfield  {author} {\bibinfo {author} {\bibfnamefont {K.~C.}\ \bibnamefont
  {Wright}}, \bibinfo {author} {\bibfnamefont {L.~S.}\ \bibnamefont {Leslie}},
  \ and\ \bibinfo {author} {\bibfnamefont {N.~P.}\ \bibnamefont {Bigelow}},\
  }\bibfield  {title} {\enquote {\bibinfo {title} {Raman coupling of zeeman
  sublevels in an alkali-metal bose-einstein condensate},}\ }\href {\doibase
  10.1103/PhysRevA.78.053412} {\bibfield  {journal} {\bibinfo  {journal} {Phys.
  Rev. A}\ }\textbf {\bibinfo {volume} {78}},\ \bibinfo {pages} {053412}
  (\bibinfo {year} {2008})}\BibitemShut {NoStop}%
\bibitem [{\citenamefont {Ke\ss{}ler}\ and\ \citenamefont
  {Marquardt}(2014)}]{Marquardt}%
  \BibitemOpen
  \bibfield  {author} {\bibinfo {author} {\bibfnamefont {Stefan}\ \bibnamefont
  {Ke\ss{}ler}}\ and\ \bibinfo {author} {\bibfnamefont {Florian}\ \bibnamefont
  {Marquardt}},\ }\bibfield  {title} {\enquote {\bibinfo {title}
  {Single-site-resolved measurement of the current statistics in optical
  lattices},}\ }\href {\doibase 10.1103/PhysRevA.89.061601} {\bibfield
  {journal} {\bibinfo  {journal} {Phys. Rev. A}\ }\textbf {\bibinfo {volume}
  {89}},\ \bibinfo {pages} {061601} (\bibinfo {year} {2014})}\BibitemShut
  {NoStop}%
\end{thebibliography}%

\end{document}